\documentclass[journal,12pt]{IEEEtran}
\ifCLASSINFOpdf
  \usepackage[pdftex]{graphicx}
\else
  \usepackage[dvips]{graphicx}
\fi
\usepackage{todonotes}

\usepackage{multirow}

\usepackage{glossaries}

\newacronym{aoa}{AoA}{Angle Of Arrival}
\newacronym{ack}{ACK}{Acknowledgement}
\newacronym{akmp}{AKMP}{Authentication and Key Management Protocol}
\newacronym{aod}{AOD}{Angle Of Departure}
\newacronym{ap}{AP}{Access Point}
\newacronym{asap}{ASAP}{As Soon As Possible}
\newacronym{awv}{AWV}{Antenna Weight Vector}
\newacronym{brp}{BRP}{Beam Refinement Phase}
\newacronym{cdf}{CDF}{Cumulative distribution function}
\newacronym{cir}{CIR}{Channel Impulse Response}
\newacronym{cli}{CLI}{Command Line Interface}
\newacronym{cots}{COTS}{Commercial off-the-shelf}
\newacronym{csi}{CSI}{Channel State Information}
\newacronym{dhss}{DHss}{Diffie-Hellman (DH) shared secret}
\newacronym{dmg}{DMG}{Directional Multi-Gigabit}
\newacronym{edca}{EDCA}{Enhanced Distributed Channel Access}
\newacronym{edmg}{EDMG}{Enhanced Directional Multi-Gigabit}
\newacronym{fpbt}{FPBT}{First Path Beamforming Training}
\newacronym{ftm}{FTM}{Fine Timing Measurement}
\newacronym{ftmr}{FTMR}{FTM Request}
\newacronym{gi}{GI}{Guard Interval}
\newacronym{hkdf}{HKDF}{HMAC-based Key Derivation Function}
\newacronym{i2r}{I2R}{Initiator-to-Responder}
\newacronym{ista}{ISTA}{Initiating Station}
\newacronym{iftm}{IFTM}{Initial FTM}
\newacronym{iftmr}{IFTMR}{Initial FTM Request}
\newacronym{ipftmr}{IPFTMR}{Initial Protected FTM Request}
\newacronym{kdf}{KDF}{Key Derivation Function}
\newacronym{lci}{LCI}{Location Configuration Information}
\newacronym{los}{LOS}{Line Of Sight}
\newacronym{mic}{MIC}{Message Integrity Code}
\newacronym{mmwave}{mmWave}{Millimiter Wave}
\newacronym{mpdu}{MPDU}{Medium Access Control (MAC) protocol data unit}
\newacronym{ngp}{NGP}{Next Generation Positioning}
\newacronym{nlos}{NLOS}{Non Line Of Sight}
\newacronym{ntb}{NTB}{Non-Triggered-Based}
\newacronym{p2p}{P2P}{Peer to Peer}
\newacronym{pasn}{PASN}{Pre-Association Security Negotiation}
\newacronym{pdp}{PDP}{Power Delay Profile}
\newacronym{pftm}{PFTM}{Protected FTM}
\newacronym{pmk}{PMK}{Pairwise Master Key}
\newacronym{pmksa}{PMKSA}{Pairwise Master Key Security Association}
\newacronym{ppdu}{PPDU}{Physical Protocol Data Unit}
\newacronym{prf}{PRF}{Pseudo-random Function}
\newacronym{prk}{PRK}{Pseudo-random Key}
\newacronym{ptb}{PTB}{Passive Triggered-Based}
\newacronym{ptk}{PTK}{Pairwise Transient Key}
\newacronym{ptksa}{PTKSA}{Pairwise Transient Key Security Association}
\newacronym{r2i}{R2I}{Responder-to-Initiator}
\newacronym{rsna}{RSNA}{Robust Security Network Association}
\newacronym{rsta}{RSTA}{Responding Station}
\newacronym{rsrp}{RSRP}{Reference Signal Received Power}
\newacronym{rssi}{RSSI}{Received Signal Strength Indicator}
\newacronym{rtt}{RTT}{Round Trip Time}
\newacronym{sta}{STA}{Station}
\newacronym{snr}{SNR}{Signal Noise Ratio}
\newacronym{soa}{SoA}{State of the Art}
\newacronym{scofdm}{SC-OFDM}{Single Carrier Orthogonal Frequency-Division Multiplexing}
\newacronym{tb}{TB}{Triggered-Based}
\newacronym{trn}{TRN}{Training}
\newacronym{tsf}{TSF}{Time Synchronization Function}
\newacronym{toa}{TOA}{Time of Arrival}
\newacronym{tod}{TOD}{Time of Departure}
\newacronym{tdoa}{TDOA}{Time-Difference-of-Arrival}
\newacronym{tof}{ToF}{Time of Flight}
\newacronym{txss}{TXSS}{Transmit sector sweep}
\newacronym{uwb}{UWB}{Ultra Wide Band}

\usepackage{lipsum}
\usepackage{enumitem}

\begin{document}
\bstctlcite{IEEEexample:BSTcontrol}
%
\title{IEEE 802.11az Indoor Positioning with mmWave}
%
%
%

\author{
Pablo Picazo-Martínez,
Carlos Barroso-Fernández,
Jorge Martín-Pérez,\\
Milan Groshev and
Antonio de la Oliva
\thanks{The authors are with the
Department of Telematics Engineering,
Universidad Carlos~III de Madrid; and
Departamento de Ingeniería de Sistemas Telemáticos,
Universidad Politécnica de Madrid.
E-mail: \{papicazo,cbarroso,mgroshev\}@pa.uc3m.es,
jorge.martin.perez@upm.es,
aoliva@it.uc3m.es.}
\thanks{This work has been partially funded by the European Union’s
Horizon Europe research and innovation programme under
grant agreement No 101095759 (Hexa-X-II),
the Spanish Ministry of Economic Affairs and Digital
Transformation and the European Union-NextGenerationEU through the
UNICO 5G I+D 6G-EDGEDT and Remote Driver (TSI-065100-2022-003), and
the Spanish Ministry of Economy and Competitiveness \& and the Spanish
Ministry of Science and Innovation through ECTICS project (PID2019-105257RB-C21).}

\thanks{©2023 IEEE. Personal use of this material is permitted. Permission
from IEEE must be obtained for all other uses, in any current or future
media, including reprinting/republishing this material for advertising
or promotional purposes, creating new or redistribution to servers or lists, or reuse of any copyrighted
component of this work in other work}
}

%
%


\markboth{ACCEPTED IN IEEE COMMUNICATIONS MAGAZINE}%
{Shell \MakeLowercase{\textit{et al.}}: Bare Demo of IEEEtran.cls for IEEE Journals}

%



\maketitle

\begin{abstract}
Last years we have witnessed the uprising of location-based applications, which depend on the device's capabilities to accurately obtain their position. IEEE 802.11, foretelling the need for such applications, started the IEEE 802.11az work on Next Generation Positioning. Although this standard provides positioning enhancements for sub-6~GHz and mmWave bands, high accuracy in the order of centimeters can only be obtained in the latter band, thanks to the high
temporal resolution from the multi-GHz bandwidth. 
This work presents the new techniques provided by IEEE 802.11az for enhanced secured positioning in the mmWave band.
Additionally, this
paper assesses 802.11az~mmWave accuracy using a novel
trigonometry solution, compares it
with advanced positioning solutions, and
identifies open research challenges.

\end{abstract}


%
\IEEEpeerreviewmaketitle
\section{Introduction}
With the advance of new services requiring precise knowledge of location such as asset tracking, factory automation, or autonomous robots; precise localization and positioning are demanded from local and personal wireless technologies~\cite{m5}. While the current sub-6\,\textrm{GHz} WLANs can achieve a positioning accuracy of tens of meters to tens of centimeters~\cite{m1}, the 60\,\textrm{GHz} WLAN with its high temporal resolution from the multi-\textrm{GHz} bandwidth can achieve centimeter positioning accuracy, not achievable in lower bands. The position accuracy can be further enhanced through the use of wider bandwidths, as currently being explored in IEEE 802.11bk task group.

Although researchers have already 
validated~\cite{mm-track,mdtrack}
the centimeter accuracy
achievable using mmWave beamforming -- see\footnote{In the paper we use blue color for legacy features, purple for extended features, and red color for new 802.11az features.}~Fig.~\ref{fig:relative-positioning},
mmWave positioning in 802.11
still faces several challenges, solved in 802.11az~\cite{802.11az}:


\begin{figure}[t]
    \centering
    \includegraphics[width=0.9\columnwidth]{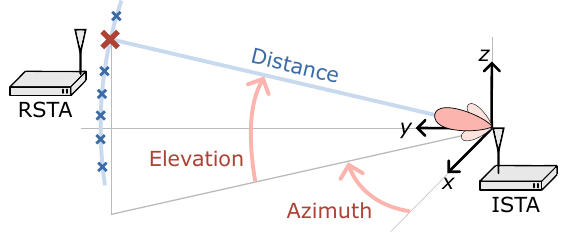}
    \caption{Legacy positioning in 802.11
    only knew the distance between devices,
    thus, there are many feasible locations
    (blue markers). 
    802.11az uses \gls{mmwave}  
    beamforming to obtain the azimuth and
    elevation between devices, hence, 
    the exact location (red marker).}
    \label{fig:relative-positioning}
\end{figure}

\begin{figure*}[t]
    \centering
    \includegraphics[width=0.9\textwidth]{%
    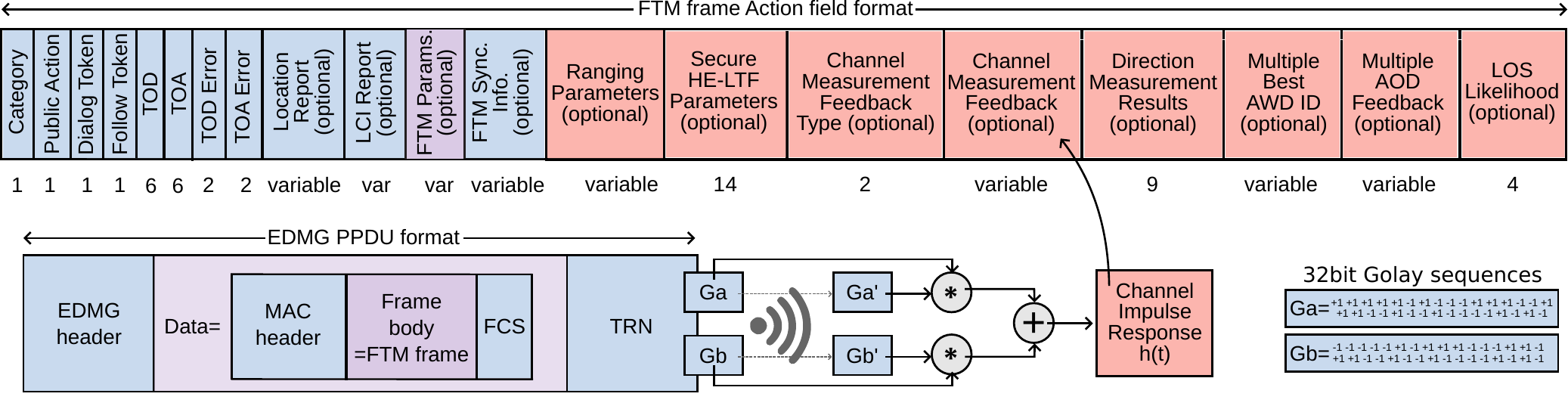}
    \caption{\gls{ftm} frames (top) are Action
    fields encapsulated inside EDMG \gls{ppdu}s
    (bottom left).
    The \gls{trn} field contains Golay sequences
    G\textsubscript{a},
    G\textsubscript{b},
    (bottom middle)
    used for Channel Measurements included
    in the \gls{ftm} frame.
    The figure shows
    \gls{ftm} legacy  fields (blue), so as 
    802.11az additions
    (red) and modifications (purple).
    Numbers below the \gls{ftm} frame
    indicate the
    number of octets that each field occupy.
    }
    
    \label{fig:ftm-frames}
\end{figure*}

\begin{enumerate}[label=(\roman*)]




\item
mmWave is highly directive and the signal path may differ in the estimation of the distance between \gls{sta}s and the angle (elevation and azimuth) of the signal (see Fig.~\ref{fig:relative-positioning}). 
To overcome such problem,
802.11az includes angle and estimations in \glsfirst{ftm} exchanges.

\item The high-frequency of mmWave signals results in increased signal attenuation, and \gls{nlos} propagation complicates the positioning
accuracy.
To enhance NLOS possitioning
accuracy,
802.11az presents 
the \gls{los} assessment
procedure.


\item
The 
positioning information is not ciphered and malicious stations may know the location of a user. This issue is solved in 802.11az by ciphering the \gls{ftm} sessions using keys derived in the \gls{pasn}.

\end{enumerate}








Recent works have studied diverse aspects covered in the 802.11az amendment. In \cite{m3}, authors study the impact of 802.11az sub-6\,\textrm{GHz} multi-user positioning on the channel delay and throughput. Authors in \cite{az-sub6-finger}~propose a deep learning solution for 802.11az positioning in sub-6\,\textrm{GHz} channels. While works as~\cite{m2} show that malicious sub-6\,\textrm{GHz} \gls{sta}s may inject wrong position reports, or attack the signal used to obtain the \gls{tof}~\cite{security-tof-ranging}.

However, the domain of 802.11az operating at mmWave still remains unexplored in the literature.
This article contributes to the \gls{soa}
($i$) explaining 802.11az mmWave positioning exchanges,
\gls{los} assessment, and security 
(Sec.~\ref{sec:ftm}-\ref{sec:secure}); and
($ii$) assessing the positioning accuracy
of future 802.11az mmWave implementations
through a novel trigonometry solution
that overcomes the \gls{nlos}
signal bounce against a wall
(Sec.~\ref{sec:experimental}).

\begin{figure}[t]
    \centering
    \includegraphics[width=.9\columnwidth]{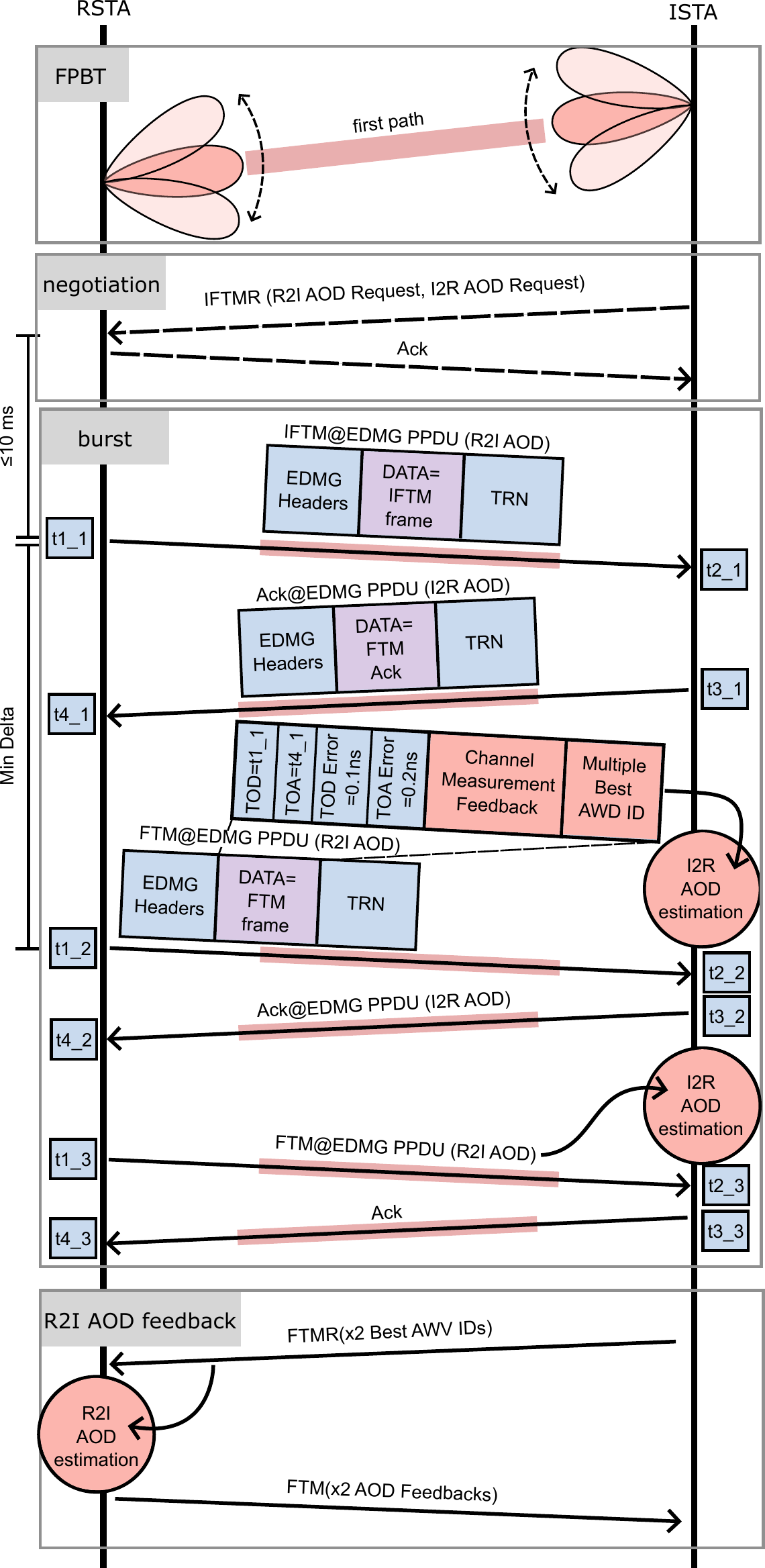}
    \caption{\gls{ftm} session over EDMG
    with \gls{fpbt},
    \gls{i2r} and \gls{r2i} AOD measurements.
    The illustration highlights legacy (blue),
    modified (purple),
    and new (red) features from
    802.11az.}
    \label{fig:ftm-exchange}
\end{figure}

\section{\glsentryshort{ftm} protocol over EDMG}
\label{sec:ftm}
802.11az exchanges positioning information using the \gls{ftm} protocol and optimizes the bandwidth utilization by sending distance and angle estimations in the same frame. 
In this section, we analyse how 802.11az
achieves Next Generation Positioning using
\gls{ftm} over
\gls{edmg}, so as
the frames encapsulation,
negotiation procedure and measurement
exchange.

\subsection{Next Generation Positioning in \gls{ftm} \& \gls{edmg}}
\label{subsec:ftm-positioning}

802.11az positioning is based on the distance, elevation and azimuth between \gls{sta}s.
While the distance is computed using the \gls{tof} of \gls{ftm} exchanges, the azimuth and elevation
angles are obtained using the \gls{mmwave} beamforming procedures defined in 802.11ay \gls{edmg}, namely short sector sweep and beam refinement. 
However,
802.11az does not specify which positioning
algorithm to use.

In \gls{edmg} beamforming the station that initiates the FTM exchange
-- the \gls{ista} -- tries different \gls{awv} configurations to point
the \gls{mmwave} beam to the \gls{rsta}.
Specifically, the \gls{ista} chooses the \gls{awv}
with the highest quality, which can be determined by the smallest \gls{tof}, highest \gls{rsrp}, \gls{rssi}, or best \gls{snr}. The method to estimate the best quality parameters is out of the scope of the standard. Once the best \gls{awv} is determined, the \gls{ista} estimates the \gls{aod} of the \gls{mmwave} beam, i.e., the departing elevation and azimuth. 


Thanks to the high bandwidth and carrier frequency of mmWave, the distance, elevation and azimuth estimations result into highly accurate relative positioning.
However, to obtain absolute positioning a \gls{lci} Report~\cite{802.11az} with latitude, longitude and altitude needs to be added inside the \gls{ftm} frame.


\subsection{\gls{ftm} encapsulation in \gls{edmg}}
\label{subsec:ftm-encapsulation}

\gls{ftm} frames are transmitted as 802.11 Action fields
-- see~Fig.~\ref{fig:ftm-frames}.
If \gls{ftm} is used between two
\gls{edmg} \gls{sta}s, the \gls{ftm} frame is sent within an \gls{edmg} \gls{ppdu}, which
allows simultaneously to
($i$) derive
channel
estimations using the
\gls{edmg} \gls{trn} field; and ($ii$) exchange
positioning estimations within the \gls{ftm}
frame.

The \gls{edmg} \gls{ppdu} contains a
\gls{trn} field with Golay sequences used
for channel estimations.
In particular, the \gls{trn} field carries
two complementary Golay sequences
G\textsubscript{a} and
G\textsubscript{b} that allow reconstructing
an accurate estimation of the \gls{cir}.
Complementary Golay sequences are binary 
sequences known by the \gls{ista}
and \gls{rsta} whose autocorrelations sum
up a delta function
-- see~\cite{golayexplain} for further details.
Consequently ($i$) correlating
the sequence received by the antenna
G\textsubscript{a}' with the original
sequence G\textsubscript{a}
(G\textsubscript{b}
with G\textsubscript{b}', respectively); and
($ii$) summing up the correlation of both
sequences results into an accurate
estimation of the \gls{cir} included within
the Channel Measurement Feedback
-- see~Fig.~\ref{fig:ftm-frames}
(bottom right).


The channel and angle estimations obtained with
\gls{trn} are carried in the 
\gls{ftm} frames. In particular,
the 802.11az amendment has
extended the \gls{ftm} frames to carry
the \gls{trn} Channel Measurements
associated to the \gls{awv},
angle estimations, and \gls{los} likelihood
-- see Fig.~\ref{fig:ftm-frames} red fields.
These new fields are complemented with
legacy timestamps, localization
and synchronization information 
-- blue color in Fig.~\ref{fig:ftm-frames} --
to obtain accurate positioning.

\subsection{Negotiating \gls{ftm} sessions
over \gls{edmg}}
\label{subsec:ftm-negotiation}

Before the \gls{ftm} negotiation,
\gls{edmg} \gls{sta}s can find the first
path among them
through the \gls{fpbt} procedure
-- see~Fig.~\ref{fig:ftm-exchange}.
Then, both negotiate
parameters as the session duration or the number of \gls{ftm} bursts within a session.

The negotiation starts with the exchange
of an \gls{iftmr} where the \gls{ista}
proposes, e.g., an \gls{edmg} \gls{ftm}
session over the first path with one burst and
\gls{aod} estimations
-- as in Fig.~\ref{fig:ftm-exchange}.
Additionally, the \gls{ista} may propose
a secure exchange using a specific bandwidth
between 2.16 and 8.64~GHz.

It is up to the \gls{rsta} to decide
whether it accepts the proposed
\gls{ftm} parameters, or it
renegotiates them.
In case the \gls{rsta}
accepts the parameters of the received
\gls{iftmr}, it sends an \gls{ack} followed
by an \gls{iftm} frame in less than 10~ms.

The negotiation fails in case the \gls{rsta} does not reply with an \gls{ack} or the \gls{iftm} frame reports that the \gls{ftm} session cannot start, i.e. \gls{rsta} is not capable of satisfying the proposed parameters.


\subsection{\gls{ftm} measurement exchange
over \gls{edmg}}
\label{subsec:measurement-exchange}

\begin{figure*}[t]
    \centering
    \includegraphics[width=0.95\textwidth]{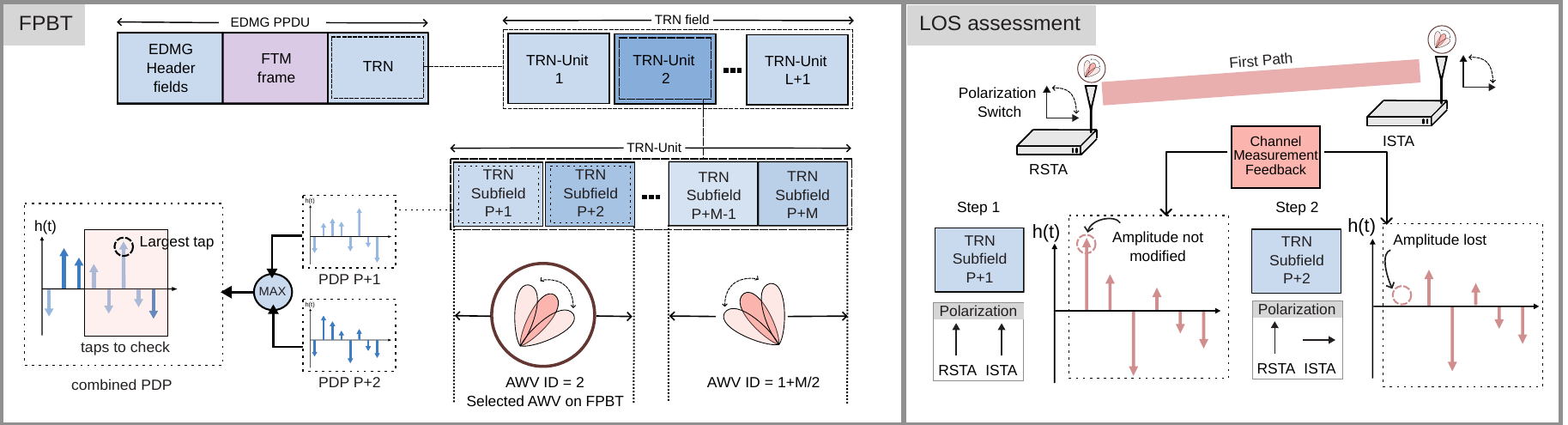}
    \caption{802.11az finds the first path
    sweeping the \gls{mmwave} beam during the
    transmission of \gls{edmg} \gls{trn}
    Subfields (left).
    Then, \gls{sta}s change
    their \gls{mmwave} antenna polarization to
    tell whether they are at \gls{los}
    by checking the Channel Measurements
    (right).}
    \label{fig:FPBT}
\end{figure*}

After a successful negotiation, \gls{sta}s
start an \gls{ftm} session to
perform distance and angle measurements.
Each session consists of bursts with multiple
exchanges of \gls{ftm} frames encapsulated
in \gls{edmg} \gls{ppdu}s.
Fig.~\ref{fig:ftm-exchange} illustrates an
\gls{ftm} session consisting of a single
burst of three \gls{ftm} and \gls{ack}
exchanges.

In every exchange the \gls{sta}s timestamp
the \gls{tod} and \gls{toa} of the sent
and received frames, respectively. As a
result, it is possible to estimate the distance
in the first exchange by multiplying the
speed of light by the
corresponding \gls{rtt} ($RTT_1$).
Fig.~\ref{fig:ftm-exchange} illustrates in
blue boxes the timestamps used in the
\gls{rtt} estimation.

However, the real time at which frames
depart and arrive may differ with the
respective timestamps. For example,
an \gls{iftm} frame
may have departed at nanosecond 1~ns,
whilst the clock timestamp may be
$t1\_1=1.1$~ns -- thus, the \gls{tod} Error of
0.1~ns shown in Fig.~\ref{fig:ftm-frames}.
Due to the speed of light, 
the timestamp error of 0.1~ns would
result in a distance estimation error of
3~cm. Hence, \gls{ftm} should obtain distance
estimations checking the exchange whose timestamps had least error.

To derive angle estimations within a burst,
the \gls{sta}s compute channel measurements
using the \gls{edmg} \gls{trn} fields.
Additionally, \gls{sta}s change their \gls{awv},
to try out multiple setups in the angle estimations.
Fig.~\ref{fig:ftm-exchange} exemplifies how 
the initiator estimates the departure angle using the channel measurements and best \gls{awv} reported by the responder in the \gls{ftm} frame-- see the
first \gls{i2r} \gls{aod} estimation in
Fig.~\ref{fig:ftm-exchange}.

In an \gls{ftm} session it is possible to
drive multiple \glsdesc{i2r} (\gls{i2r}) \gls{aod} estimations
during the burst -- two estimations in
Fig.~\ref{fig:ftm-exchange} example. But in
the case of \glsdesc{r2i} (\gls{r2i}) \gls{aod},
802.11az specifies that
the estimations are done once the burst
finishes. Specifically, the initiator
exchanges the best \gls{awv} setups perceived
during the burst so the responder can estimate
its \gls{aod}
-- see~Fig.~\ref{fig:ftm-exchange}. Finally,
the responder sends back to the initiator
the two angle estimations.

\section{\gls{los} assessment over first path}
\label{sec:fpbtlos}





802.11az uses \gls{fpbt} to determine the best path between two \gls{sta}s and \gls{los} assessment to determine if that path suffers or not from reflections.
Both \gls{fpbt} and \gls{los} are essential procedures
for high accuracy in \gls{mmwave}.

\subsection{\glsdesc{fpbt} (\gls{fpbt})}
In the \gls{fpbt} procedure, the \gls{sta}s
try different \gls{mmwave} antenna
configurations until the beam points to the
first path. In particular, the \gls{sta}s sweep 
over multiple \gls{awv} configurations and
select the best one
-- see~Fig.~\ref{fig:FPBT} (left).
In the following, we detail the \gls{fpbt}
procedure presenting
($i$) the \gls{edmg} \gls{trn} fields; and
($ii$) the \gls{awv} sweep procedure.


The \gls{edmg} \gls{ppdu}s contains the \gls{trn} field that is used to perform
channel estimations in the \gls{fpbt} procedure.
The \gls{trn} field contains L+1
\gls{trn}-Units 
-- see~Fig.~\ref{fig:FPBT}~(left).
Every \gls{trn}-Unit contains P+M \gls{trn}
Subfields that are filled with Golay sequences.
The transmission of the Golay sequence 
results in a \gls{pdp}
that measures the in-phase, quadrature and
\gls{snr} of each tap -- see the \gls{pdp}s of \gls{trn} Subfields P+1 and P+2 in Fig.~\ref{fig:FPBT}~(left).

The \gls{awv} sweep procedure 
changes the \gls{awv} after the transmission of
a group of \gls{trn} Subfields
-- e.g., after a group of two in
Fig.~\ref{fig:FPBT}~(left). 
Then, the \gls{sta} combines the \gls{pdp}s
of each \gls{trn} Subfield in the group, and
selects the highest amplitude taps.
The quality of the combined \gls{pdp} is
measured by checking the taps near the
one with highest amplitude
-- see~Fig.~\ref{fig:FPBT}~(left).
Finally, the \gls{sta} compares the combined
\gls{pdp}s in the \gls{trn} field,
and selects the best \gls{awv} --
which points to the first path.

\subsection{\glsdesc{los} (\gls{los}) Assessment }
\label{subsec:los-assessment}
Once \gls{fpbt} determines the best \gls{awv}, the \gls{ista} initiates an \gls{ftm}
\gls{los} assessment exchange. In the exchange
the \gls{sta}s alternate the antenna
polarization to determine whether the
first path is at \gls{los}.
The change of polarization is perceived
in the taps of the received signal,
whose values are reported as a list of in-phase and
quadrature (I/Q) components within the
Channel Measurement Feedback.

The \gls{los} assessment compares the
signal taps of the received \gls{trn} Subfields
P+1 and P+2. The \gls{trn} Subfield P+1
is transmitted/received by \gls{sta}s using
the same polarization.
\gls{trn} Subfield P+2
is transmitted/received using vertical/horizontal polarization
-- see~Fig.~\ref{fig:FPBT}~(right). If \gls{sta}s are
at \gls{los}, the main tap amplitude
for \gls{trn} Subfield P+1 is non-zero, and
zero for \gls{trn} Subfield P+2 due to polarization
mismatch. Upon \gls{nlos}, the main tap amplitude
is not zero for both \gls{trn} Subfields P+1 and
P+2 because reflections change the
polarization.

Once the procedure is completed, \gls{sta}s elaborate a \gls{los} likelihood report with the probability of being at \gls{los}/\gls{nlos}.
How to compute the \gls{los} likelihood is
out of the scope of 802.11az, however,
works as~\cite{losnlos} propose alternatives to distinguish between \gls{los}/\gls{nlos} even without antenna alignment.
Furthermore, if the \gls{sta} is at \gls{nlos}, the
position is obtained considering the
reflection of walls -- as in the
trigonometry solution we propose in
Sec.~\ref{subsec:accuracy}.

\section{Secure next-generation positioning}
\label{sec:secure}

The position of a \gls{sta} is sensitive information that malicious \gls{sta}s may sniff or corrupt~\cite{m2}.
With 802.11az it is possible to achieve secure \gls{ftm} sessions.
Specifically, \gls{sta}s authenticate
through \gls{pasn}
to later cipher the exchanged \gls{ftm} frames.


\subsection{\glsdesc{pasn} (\gls{pasn})}
\label{subsec:pasn}




802.11az introduces \gls{pasn} for secure exchanges among non-associated \gls{sta}s. \gls{pasn} reaches a \gls{ptksa} between \gls{ap} and \gls{sta} {acting as a \gls{rsna} protocol.} However \gls{pasn} can also reach the \gls{ptksa} without the \gls{pmk}, acting as a non-\gls{rsna} protocol.
%

The \gls{ptksa} key is obtained exchanging
three \gls{pasn} elements
(inside authentication frames) -- see Fig.~\ref{fig:security}.
First, the \gls{sta} sends to
the \gls{ap} an ephemeral key and
parameters to later derive a \gls{pmk}.
Second, the \gls{ap} sends to the \gls{sta}
a protected frame with its ephemeral key and the chosen parameters
to derive the \gls{pmk}.
Third, the \gls{sta} confirms the chosen
parameters sending an \gls{ack}
within a protected frame.

Upon confirmation, the \gls{sta} and
\gls{ap} use the \gls{kdf} to
generate the \gls{ptksa} key using
the parameters negotiated in the
exchanged \gls{pasn} frames.

\subsection{Secure \gls{ftm} sessions}
\label{subsec:secure-ftm}
\begin{figure}[t]
    \centering
    \includegraphics[width=.85\columnwidth]{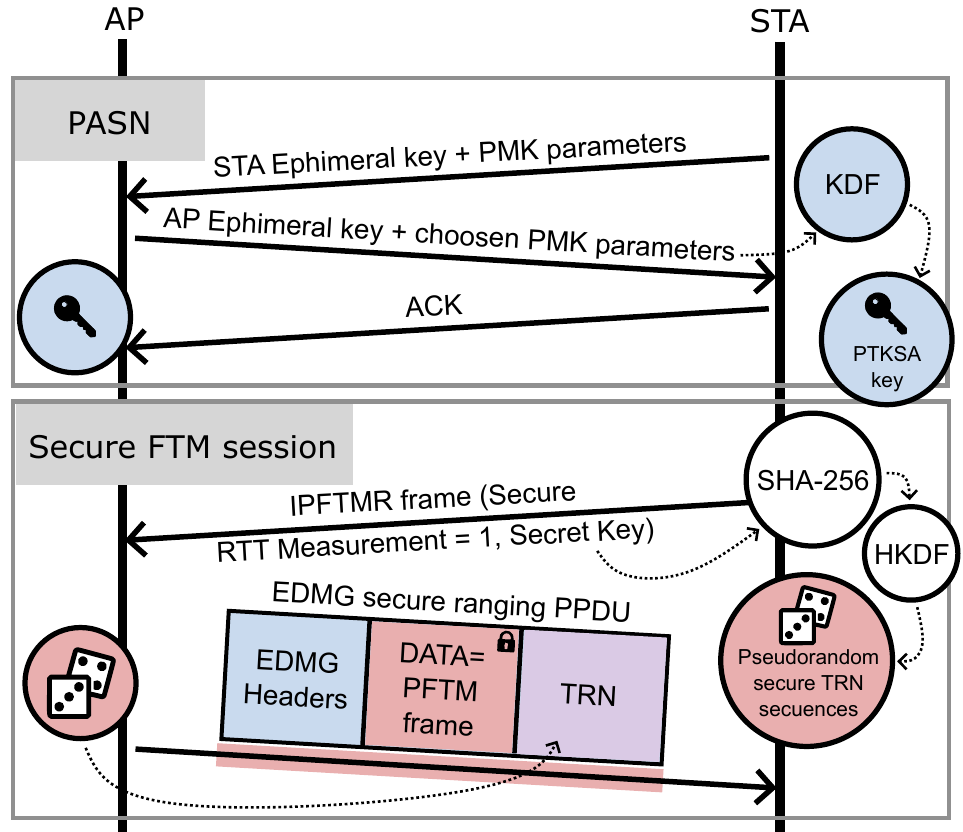}
    \caption{
        The \gls{sta} and \gls{ap}
        securely pre-associate using
        802.11az \gls{pasn} (top).
        Afterward, both protect the
        \gls{ftm} session (bottom)
        ciphering the \gls{ftm} frame
        (black lock), and generating
        pseudorandom secure
        \gls{trn} sequences
        at PHY level (dice).
    }
    \label{fig:security}
\end{figure}



Once the \gls{ptksa} is achieved
through \gls{pasn}, or legacy authentication protocols~\cite{802.11az},
the \gls{sta} requests a Secure
\gls{ftm} session issuing an \gls{ipftmr}
-- see~Fig.~\ref{fig:security}. 
Then, the \gls{sta} obtains its relative position with respect to the \gls{ap} 
through \gls{edmg} secure ranging \gls{ppdu}s. 
The exchanged \gls{ppdu}s contain
\gls{pftm} frames and
secure \gls{trn} Subfields for channel and
positioning estimations.

The \gls{pftm} frame is ciphered using
a \gls{rsna} or non-\gls{rsna} 802.11 confidentiality and integrity protocol~\cite{802.11az}.
Hence, 
the \gls{rsta} will notice
if a malicious \gls{sta}
alters/corrupts the positioning information.

Secure \gls{trn} Subfields consist of
padded pseudorandom bit sequences transmitted
using \mbox{$\tfrac{\pi}{2}$-BPSK}. The pseudorandom
sequences are generated as follows.
First, the \gls{sta} creates a
\gls{prk} using an SHA-256 hash function
that receives the Secret Key and the
identifier of the \gls{pmk} previously
generated, e.g., in \gls{pasn}. Second,
the \gls{sta} produces the pseudorandom
secure \gls{trn}
sequence by feeding the \gls{hkdf}
with ($i$) the \gls{prk};
($ii$) the string ''EDMG Secure RTT''; and
($iii$) the length of the \gls{trn} field. 
A 3\textsuperscript{rd}
malicious \gls{sta} sniffing the secure \gls{trn} Subfields cannot estimate the \gls{cir} and derive other STA position, as it does not know the original pseudorandom sequence.
Moreover, \mbox{$\tfrac{\pi}{2}$-BPSK} is robust
enough to
detect if the TRN Subfield is corrupted
upon channel jamming, and
discard such TRN Subfield
for positioning estimations.

\begin{figure}[t]
    \centering
    \includegraphics[width=.9\columnwidth]{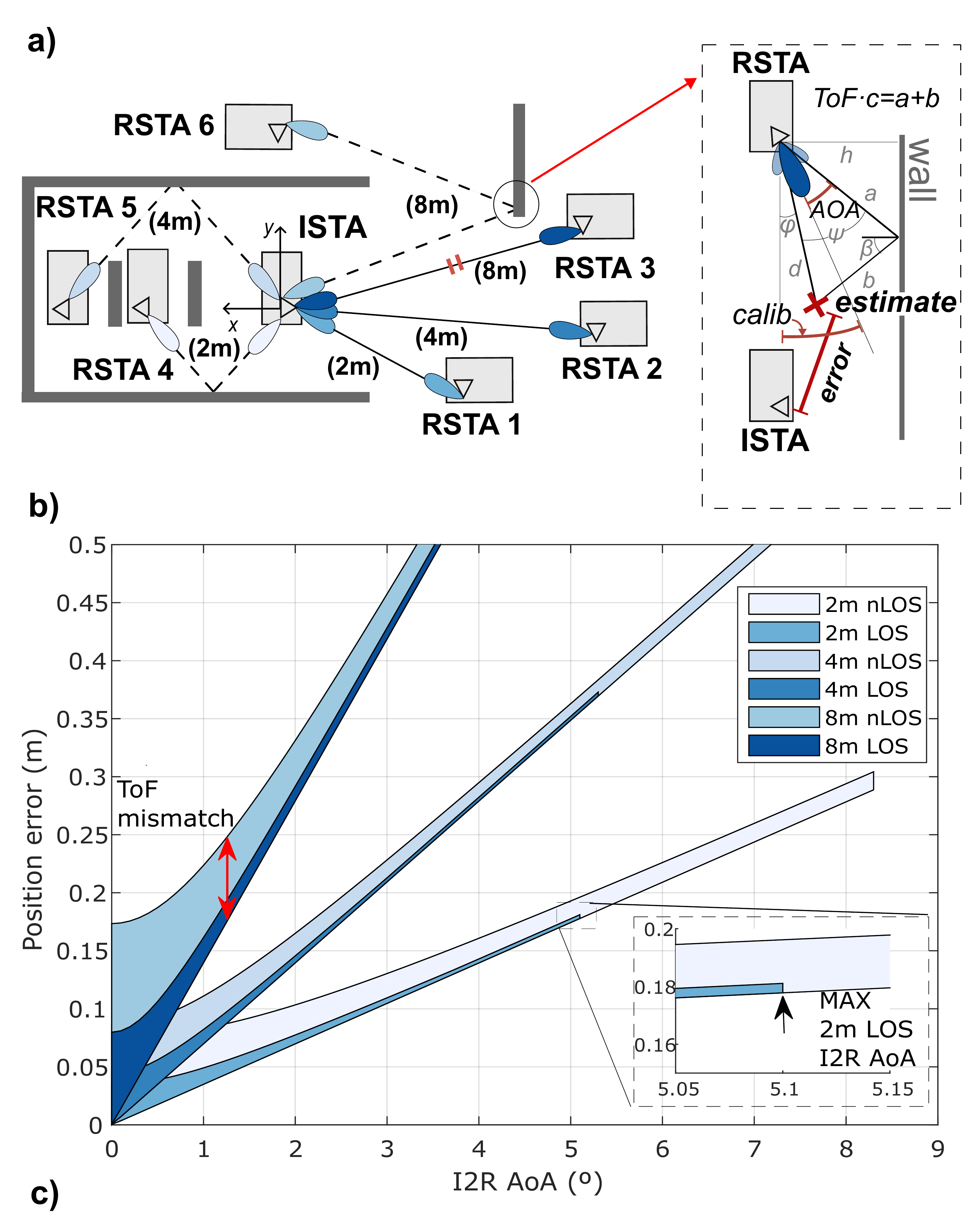}

\resizebox{.9\columnwidth}{!}{%
\begin{tabular}{|c|l|llll|}
\hline
\multirow{2}{*}{Distance} & \multicolumn{1}{c|}{\multirow{2}{*}{Technology comparison}} & \multicolumn{4}{c|}{Percentile Error (cm)}                      \\ \cline{3-6} 
                          & \multicolumn{1}{c|}{}                            & 25 \%           & 50 \%           & 75 \%            & 100 \%           \\ \hline
\multirow{2}{*}{7m}       & 3GPP Sub-6\,GHz \cite{3gpp1}                                   & 2.00          & 8.25          & 15.50          & \textbf{30.00} \\
                          & 802.11az mmWave                                         & \textbf{1.76} & \textbf{3.13} & \textbf{5.02}  & 33.15          \\ \hline
\multirow{2}{*}{7.07m}    & Bluetooth 5.1 \cite{bluetooth}                                    & 30.00         & 37.50         & 46.50          & 80.00          \\
                          & 802.11az mmWave                                         & \textbf{4.99} & \textbf{5.85} & \textbf{11.46} & \textbf{36.49} \\ \hline
\multirow{2}{*}{9m}       & 3GPP Sub-6\,GHz  \cite{3gpp1}                                  & \textbf{3.00} & 8.00          & 16.00          & 56.00          \\
                          & 802.11az mmWave                                         & 6.19          & \textbf{7.72} & \textbf{15.65} & \textbf{33.67} \\ \hline
\multirow{2}{*}{11.2m}    & 3GPP mmWave \cite{3gppmmwave}                                     & 6.50          & \textbf{7.80} & \textbf{15.00} & 80.00          \\
                          & 802.11az mmWave                                         & \textbf{5.67} & 7.91          & 20.35          & \textbf{52.75} \\ \hline
\multirow{2}{*}{14.2m}    & 3GPP mmWave \cite{3gppmmwave}                                      & 7.50          & 10.50         & \textbf{18.75} & 85.50          \\
                          & 802.11az mmWave                                         & \textbf{4.39} & \textbf{9.78} & 22.34          & \textbf{56.30} \\ \hline
\end{tabular}
}

    \caption{Experimental scenario with trigonometry approach in \gls{nlos} (a). \gls{i2r}
    \gls{aoa} errors vs. the
    positioning error (b).
    Legacy \gls{ftm} does not use
    \gls{fpbt}, introducing positioning
    error variance (\gls{tof} mismatch). Comparison between \gls{soa} alternatives for indoor positioning (c).}
    \label{fig:results}
\end{figure}

\section{{Evaluation of 802.11az mmWave}}
\label{sec:experimental}
{This section evaluates the accuracy of 802.11az implementations using a novel trigonometry
solution\footnote{{https://gitlab.netcom.it.uc3m.es/papicazo/802.11az/}}. Additionally, we compare 802.11az performance
with other indoor positioning solutions.}

\subsection{{Positioning accuracy}}
\label{subsec:accuracy}



{In our experimental analysis, we evaluate the precision of 802.11az implementations. For this purpose, we use MikroTik AP 60G, commercial-off-the-shelf devices with legacy 802.11-2016~\gls{ftm} support and a \gls{mmwave} {uniform rectangular} antenna array of 36 elements, using 2.16~GHz channel bandwidth, as specified in 802.11az. 
To obtain the \gls{tof} and \gls{i2r} \gls{aoa} estimations -- 3\textsuperscript{rd} option of~\cite[Table~11.11]{802.11az}) --  we apply, respectively: ($i$) 802.11ad legacy \gls{ftm}; and ($ii$) beamsweeping using a custom version of OpenWrt and mD-Track~\cite{mdtrack} to classify the obtained paths, extracting
the \gls{aoa} (elevation \& azimuth) and received power.}

{The scenario consists of an \gls{ista} and six \gls{rsta}s, half at \gls{los}, half at \gls{nlos}, arranged in the room illustrated in Fig.~\ref{fig:results}\,(a). The three \gls{rsta}s with \gls{los} are placed 2, 4, and 8\,m away from the \gls{ista}; and the signals of \gls{nlos} \gls{rsta}s travel 2, 4 and 8\,m to reach the \gls{ista}. All the \gls{sta}s are placed 1\,m above the floor, thus the beam's elevation is 0º.

When \gls{sta}s are at \gls{los},
our trigonometry solution\footnotemark[2]
computes the relative position
leveraging the \gls{aoa} and
the \gls{tof}
estimation.
In the \gls{nlos} case
(attenuation around 5\,\textrm{dB}), our solution
obtains the triangle formed by the
signal bounce against the wall
-- with sides $a,b$ and
angle $\psi$ in
Fig.~\ref{fig:results}\,(a) zoom.
Sides $a,b$ sum up
the \gls{tof} times the speed of
light, and angle $\psi$ is obtained
through elementary trigonometry.

}


{Fig.~\ref{fig:results}\,(b) plots the position error (y-axis) depending on the \gls{i2r} \gls{aoa} error (x-axis) 
for each \gls{rsta}s in Fig.~\ref{fig:results}\,(a). As the graph shows, the \gls{rsta}s with \gls{los} experience less \gls{aoa} error (dark blue areas) than the \gls{nlos} (light blue areas) \gls{rsta}s. Fig.~\ref{fig:results}\,(b) inset evidences the latter, for the \gls{rsta} at 2\,m~\gls{los} has a maximum \gls{aoa} error of 5.1º in while the \gls{rsta} at 2\,m~\gls{nlos} achieves an 8.3º \gls{aoa} error.}


{In addition, for a fixed \gls{i2r} \gls{aoa} error, the \gls{rsta} reports different positioning errors -- see Fig.~\ref{fig:results}~(b) red arrows. This happens because the implementation does not carry \gls{aoa} estimations though the \gls{ftm} frame, where \gls{tof} is carried. This leads to an \gls{tof} mismatch with the \gls{aoa} estimation, which introduces variance in the position error, as the channel may not be the same during both estimations. Future implementations of 802.11az will overcome this problem, including the \gls{aoa} estimations in the \gls{ftm} frame.}

\subsection{{Comparison with existing solutions}}


{We compare the positioning accuracy of our implementation
with the following \gls{soa} indoor positioning technologies:
3GPP sub-6 GHz \cite{3gpp1}, Bluetooth 5.1 \cite{bluetooth},
and 3GPP mmWave \cite{3gppmmwave}.
Specifically, we compare the accuracy of our implementation
in the same scenarios where the technologies were evaluated,
each with the distances in~Fig.~\ref{fig:results}\,(c).





In the Bluetooth 5.1 scenario,
with \gls{sta}s 7.07\,m away,
802.11az~mmWave has errors below
36.49\,cm, while
Bluetooth 5.1 experiences
larger errors of up to
80\,cm due to low power and carrier.

For the 3GPP sub-6\,GHz scenarios,
with \gls{sta}s 7 and 9\,m away, we also see
larger positioning errors than in 802.11az~mmWave
due to the smaller carrier frequency.
3GPP sub-6\,GHz only outperforms our implementation
in two percentiles (7\,m 100\% and 9\,m 25\%) because of bad
\gls{aoa} estimations.

Lastly, in the 3GPP mmWave scenarios,
at 11.2 and 14.2\,m, we observe that our
implementation achieves higher worst-case accuracy
-- always below 56.30\,cm error compared to
the 85.50\,cm error of 3GPP mmWave. Although both technologies
show high accuracy, 802.11az mmWave (2.16\,GHz at @60\,GHz carrier) outperforms 3GPP mmWave (400\,MHz at @28\,GHz carrier) as it uses higher
frequency and bandwidth.}

\section{{Conclusion and Open Challenges}}
\label{sec:conclusion}

{
This work presents the main features
of 802.11az mmWave positioning. The new amendment:
($i$) enhances \gls{ftm} to include angle and
\gls{tof} estimations;
($ii$) introduces a \gls{los} assessment
procedure; and ($iii$) secures \gls{ftm}
exchanges through \gls{pasn}.
Experimental results show cm-level
accuracy using a novel trigonometry approach,
and competitive errors with respect
to \gls{soa} technologies.


Future evolutions of
802.11az~mmWave will have to tackle
the following research challenges
to enhance indoor positioning:
($i$) how to compute the \gls{los} likelihood in
the \gls{los} assessment;
($ii$) how to preserve positioning accuracy
in secure \gls{ftm} sessions
 -- e.g. using Golay sequences ciphered with
homeomorphisms~\cite{homeomorphism};
($iii$) how to coordinate 802.11az with
other technologies for multi-RAT positioning; and
($iv$) how to optimally schedule
\gls{ftm} sessions to not detriment
the wireless communication;
}


%



\ifCLASSOPTIONcaptionsoff
  \newpage
\fi





\bibliographystyle{IEEEtran}
\bibliography{IEEEabrv,IEEEsettings,bare_jrnl}{}

%
%
%

%

\vspace{-3em}
\begin{IEEEbiographynophoto}{Pablo Picazo-Martínez}
got his M.Sc. in 2022 and is a Ph.D. student at Universidad Carlos III de Madrid.
\end{IEEEbiographynophoto}
\vspace{-4em}
\begin{IEEEbiographynophoto}{Carlos Barroso-Fernández}
got his M.Sc. in 2022 and is a Ph.D. student at Universidad Carlos III de Madrid.
\end{IEEEbiographynophoto}
\vspace{-4em}
\begin{IEEEbiographynophoto}{Jorge Martín-Pérez}
got his M.Sc. in 2017 and Ph.D. in 2021,
is an assistant professor 
at Universidad Politécnica de Madrid.
\end{IEEEbiographynophoto}
\vspace{-4em}
\begin{IEEEbiographynophoto}{Milan Groshev}
got his Ph.D. on Telematics Engineering in 2022
at Universidad Carlos III de Madrid, where he works
as postdoc.
\end{IEEEbiographynophoto}
\vspace{-4em}
\begin{IEEEbiographynophoto}{Antonio de la Oliva}
 got his M.Sc. in 2004 and his Ph.D. in 2008,
is an associate professor at Universidad Carlos III de Madrid.
\end{IEEEbiographynophoto}







\end{document}